\documentclass{article}

\usepackage{arxiv}
\usepackage[numbers]{natbib}
\usepackage{afterpage}
\usepackage{epsfig}

\usepackage{amssymb} 
\usepackage{amsmath} 
\usepackage{amsfonts} 
\usepackage{amsthm} 
\usepackage{empheq} 
\usepackage{mathabx} 
\usepackage{fixmath}
\usepackage{siunitx}
\usepackage{xfrac}
\usepackage{multirow}

\usepackage{color,soul}

\usepackage[table,xcdraw]{xcolor}
\usepackage[normalem]{ulem}

\usepackage{url}

\title{Explainable Artificial Intelligence driven mask design for self-supervised seismic denoising}

\author{
  Claire Birnie \\
  KAUST\\
  Thuwal, Kingdom of Saudi Arabia \\
  \texttt{claire.birnie@kaust.edu.sa}
   \And
  Matteo Ravasi \\
  KAUST\\
  Thuwal, Kingdom of Saudi Arabia \\
  \texttt{matteo.ravasi@kaust.edu.sa}
    }
    
\begin{document}

\chead{XAI for self-supervised seismic denoising}

\maketitle

\begin{abstract}
The presence of coherent noise in seismic data leads to errors and uncertainties, and as such it is paramount to suppress noise as early and efficiently as possible. Self-supervised denoising circumvents the common requirement of deep learning procedures of having noisy-clean training pairs. However, self-supervised coherent noise suppression methods require extensive knowledge of the noise statistics. We propose the use of explainable artificial intelligence approaches to “see inside the black box” that is the denoising network and use the gained knowledge to replace the need for any prior knowledge of the noise itself. This is achieved in practice by leveraging bias-free networks and the direct linear link between input and output provided by the associated Jacobian matrix; we show that a simple averaging of the Jacobian contributions over a number of randomly selected input pixels, provides an indication of the most effective mask to suppress noise present in the data. The proposed method therefore becomes a fully automated denoising procedure requiring no clean training labels or prior knowledge. Realistic synthetic examples with noise signals of varying complexities, ranging from simple time-correlated noise to complex pseudo rig noise propagating at the velocity of the ocean, are used to validate the proposed approach. Its automated nature is highlighted further by an application to two field datasets. Without any substantial pre-processing or any knowledge of the acquisition environment, the automatically identified blind-masks are shown to perform well in suppressing both trace-wise noise in common shot gathers from the Volve marine dataset and colored noise in post stack seismic images from a land seismic survey.
\end{abstract}



\section{Introduction}
Machine learning, and in particular Deep Learning (DL), has recently become part of everyone's life; geophysics has not been immune to such a trend, and as geophysicists, we have experienced the great benefits it can bring in terms of quality and speed of execution. In the last decade, a swath of new algorithms have been proposed across the entire seismic processing pipeline that leverage DL-based methodologies. As deep learning is particularly powerful in finding patterns in complex, high-dimensional data, it has been successfully adopted for noise suppression/signal enhancement tasks across many domains, ranging from natural images \cite[e.g.,][]{Li2021} to medical images \cite[e.g.,][]{Suzuki2017} to synthetic-aperture radar images \cite[e.g.,][]{Mohan2021}. Within geophysics, different neural network architectures have been proposed for the suppression of a variety of noise signals. For example, the Denoising CNN (DnCNN) architecture has been used for random noise suppression \cite{Zhao2018}, a convolutional auto-encoder for erratic trace-wise noise suppression \cite{Qian2022}, and generative adversarial networks for ground-roll suppression \cite{Kaur2020}, just to name a few. In addition to this, a number of studies have been focused on developing networks that can tackle the complex noise field as a whole \cite[e.g.,][]{Dong2021,Zhong2021,Birnie2022}.

The majority of the denoising networks proposed so far are trained in a supervised manner, requiring pairs of noisy-clean training data, often obtained by either pre-processing of the noisy dataset by means of conventional (non-DL) denoising procedures or through the creation of realistic synthetic training data that is highly-similar to the field application data. Given that creating training data in such a manner is far from trivial, several researchers have turned their interest towards the application of self-supervised procedures for denoising tasks. Here, the noisy data represents both the input to the network and the network's training target, often with the input undergoing some minor modifications such that the input and target are no longer identical. Arguably, AutoEncoders (AEs) were the first of such approaches, where the input is reduced (encoded) to a latent space representation of a significantly smaller dimensionality prior to being up-scaled (decoded) back to the original space. Assuming the seismic signal of interest is the dominant component in the data, and a suitably small latent space is used, AEs have been shown to successfully suppress random noise in seismic data \cite{Saad2020}. Similarly, AEs can be used to attenuate coherent noise, such as high-frequency noise components in marine seismic data \cite{Hamidi2020} or diffractions from stacked data \cite{Markovic2022}. Blind-spot networks represent another family of self-supervised methods recently introduced in computer vision: to counter the challenge of sourcing noisy-clean training pairs, \cite{Krull2019} proposed a pre-processing methodology to `blind' a central pixel from the network's input, which they termed Noise2Void (N2V). Under the assumption that noise is random and independent among neighbouring pixels, and due to the removal of information from the central pixel, the network can only learn to recreate the coherent signal - hence, its ability to act as a natural denoiser. \cite{Birnie2021} illustrated the potential of blind-spot networks for suppression of random and pseudo-random noise from seismic data. Acknowledging that noise is rarely random, \cite{Broaddus2020} extended the methodology of blind-spot networks to create a blind-mask network where all pixels within a masked area have their values replaced; thereby, preventing them from contributing to a central pixel's predicted value. \cite{Broaddus2020} showed that when the mask covers noise with a high correlation to the central pixel, their blind-mask methodology, termed StructN2V, could efficiently suppress spatially-structured noise in fluorescence microscopy images.

One of the biggest challenges with StructN2V is selecting a mask that blinds all pixels that can `leak' noise information into the network's input, whilst retaining enough pixels containing signal that the network can use them to reproduce the signal in the central pixel. In seismic applications, \cite{Liu2022} used prior knowledge of the noise's spatio-temporal characteristics to design a trace-wise noise mask. Similarly, \cite{Wang2021} and \cite{Luiken2022} adapted the trace-blinding procedure for seismic deblending purposes. Whilst, \cite{Mosser2022a} proposed an extended mask for the StructN2V workflow for the suppression of dipping and hyperbolic noise instances. In all applications, the purpose of the mask is to ensure that neighbouring pixels contributing to a central pixel's predicted value do not contain coherent noise, as this could be detrimental to the network's training process. In order to analyse the influence of neighbouring pixels, we propose to adapt a methodology from the field of eXplainable Artificial Intelligence (XAI). More precisely, XAI is used here  to remove the black-box element of machine learning and provide meaning to a prediction by highlighting how the input has been utilised to compute such an output. To date, there has been relatively little use of XAI within geophysics. \cite{Benkert2021} and \cite{Noh2023} both utilise XAI methods for analysing NN predictions in facies classification, whilst \cite{Chamorro2023} investigates the contribution of different areas in a seismic shot gather with respect to the prediction of the associated surface wave dispersion curves.

In this study, we propose a workflow that leverages self-supervised, blind-spot networks in combination with XAI for the identification and subsequent application of an optimal noise mask for the training of blind-mask networks; this ultimately results in an automated denoising scheme for coherent noise suppression, which requires neither a labelled training dataset nor prior knowledge of the statistical properties of the noise. The proposed workflow is initially validated on a synthetic data for the scenarios of colored noise, trace-wise noise and pseudo-rig noise. Subsequently,  the workflow is further tested on  the Volve field dataset to denoise trace-wise noise present in the common shot gathers, before being utilised for psuedo-random noise suppression of a post-stack image from a land seismic survey. The successful suppression of a wide range of noise types, as well as the application to data at varying stages of processing, highlights the robustness of the proposed procedure.

\section{Methodology}
This study combines self-supervised blind-spot networks as presented in \cite{Krull2019} with \cite{Mohan2021}'s proposal of using the Jacobian matrix of the network to study the behaviour of bias-free neural networks. In this section, we detail these two core concepts before outlining how we combine them to determine an optimal denoising mask for the suppression of coherent noise in seismic data.

\subsection{Self-supervised, blind-mask denoisers}
Similar to statistical methods like local or non-local means \cite{Buades2005}, blind-spot denoising aims to reproduce a central pixel's value using only the values of neighbouring pixels. In this paper, we use a variation of the original (N2V) blind-spot methodology of \cite{Krull2019}. First, a number of pixels that will become the focus of the training procedure are selected; throughout this work we will refer to these pixels as \textit{active pixels}. To create the so-called blind-spots in the input image, the original value of each active pixel is replaced by the value of a randomly selected pixel within a radius, $r$. This selection and corruption process is repeated at every epoch, prior to passing the corrupted image to the network; thereby, the location and values of the active pixels change on every exposure to the network. As a self-supervised procedure, the network's target is the raw noisy data. Due to the fact that the network focuses on learning \textit{only} from neighbouring pixels, its performance is only evaluated at the active pixel locations, i.e., at the blind-spots. This is implemented by computing the training loss (for example, the $L_1$ norm) across the full image before applying a binary mask such that only the values at the active pixels are effectively evaluated. As with most deep learning procedures, the prediction errors are then back-propagated through the network and used to update its parameters to be used when evaluating the network in the next training epoch.

As previously mentioned, the blind-spot approach is only applicable for the scenario of random noise -- i.e., noise that is independent between pixels. In order to tackle noise that presents any degree of coherency, a \textit{blind-mask} is required instead of a single \textit{blind-spot} - where the mask shape is designed to remove the influence of any pixels that contain noise correlated to that within the central (active) pixel. Similar to the original procedure, a number of active pixels are initially selected. In this instance, not only is the active pixel's value replaced, as are the values of all pixels within the masked area, where the mask is centered on the active pixel. Due to the larger number of pixels to be replaced, the replaced values are now extracted from a pre-selected noise distribution, as opposed to using a value from a neighbouring pixel. The corrupted image is then passed through the network as the training input, the raw image is used as the training target, and the loss is calculated only at the active pixel locations. 

It is important to note that the implementation of the blind-spots (or masks) is only performed at the training stage. During inference, the raw, noisy images are fed directly to the trained network, without any need for pre-processing or masking of the predictions. Finally, it is worth remembering that since blind-spot networks are convolutional neural networks, they benefit from the ability convolutions have that allow them to be applied to inputs of varying shapes/sizes. As such, whilst training may be performed using patches of the training data, the inference process is usually carried out on the entire data to be denoised.

\subsection{Input Contribution Analysis via the Jacobian Matrix}
Feed-forward neural networks with REctified Linear Units (ReLUs) activation functions, such as the commonly-used UNet architecture, are piecewise affine transformations: as a consequence, one can express the nonlinear transformation $y=f(x)$ as the sum of two terms~\cite{Mohan2021}
\begin{equation}
y = f(x) = Jx + b
\end{equation}
where $x$ and $y$ are the input and output of size $N_{h} \times N_{w}$, $J$ is the Jacobian of size $N_{h}N_{w} \times N_{h}N_{w}$ of $f(\cdot)$, and $b$ is the net bias of the same size of the input. In the special case of bias-free networks (i.e., networks whose additive constants are all fixed to zero), the net bias disappears, and the complex, non-linear relationship between input and output resorts to the dot product between the Jacobian and the input. Note that the relationship between the input and output is only \textit{locally-linear}; in other words, it is only linear between one given input and its corresponding output. As such, the benefit of non-linearity introduced by the use of neural networks remains. \cite{Mohan2021} used this principle to study the behaviour of a bias-free, convolutional supervised denoiser.

To be able to apply the same concept to self-supervised denoisers, we utilise here a bias-free UNet architecture; one can therefore consider the denoised product as the application of an adaptive linear transform where each pixel in the denoised image is a weighted sum of the input noisy pixels. Note that this is nothing more than the non-local means algorithm, with the main difference that the weights are learned via a neural network instead of being computed using simple pre-defined functions. 

Evaluated during any stage of training or at the inference stage, the Jacobian matrix identifies the weighting of any given pixel from the input image on the prediction of a selected pixel in the denoised image. In this sense, the Jacobian matrix can be utilised in two manners for the analysis of blind-spot networks: 
\begin{itemize}
    \item firstly, during the training stage the computation of the Jacobian matrix can highlight whether the network is learning to avoid utilizing the blind-spots; in other words, utilising the Jacobian matrix we can compute the influence that a blind-spot pixel's value has on its own prediction. In theory, this influence should tend to zero during the training process; and,
    \item secondly, the Jacobian matrix can also indicate the neighbouring pixels which heavily influence the the blind-spot's predicted value. 
\end{itemize}

In this study, we leverage the second opportunity and utilise the Jacobian matrix to gain knowledge of the neighbouring pixels' contributions to a central pixel's predicted value. In the instance of the application of a blind-spot network on data contaminated with coherent noise, the neighbouring pixels' influence will highlight not only the pixels that contribute to the recreation of the central pixel's signal value, but also the pixels that contribute to the recreation of the central pixel's noise value - where we consider the central pixel's raw value as a summation of a signal component and noise component. Therefore, by computing the Jacobian matrix over a large number of pixels at the inference stage of a blind-spot network, a noise mask can be designed to evade the contribution of pixels that are identified as high influencers for later use with a self-supervised, blind-mask denoiser.

\begin{figure}[!htb]
  \centering
  \includegraphics[width=0.9\textwidth]{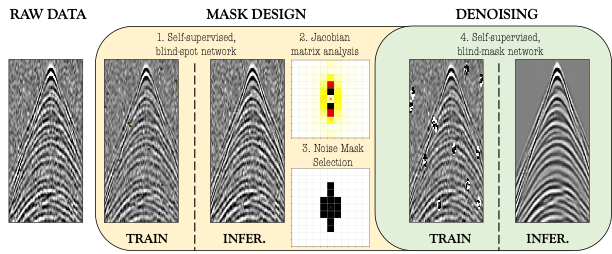}
  \caption{Proposed denoising workflow that involves: 1) performing a standard, self-supervised, blind-spot denoising approach; 2) identifying the contributing pixels through a Jacobian matrix analysis; 3) designing a noise mask that removes contributing pixels from the training input; and, 4) training a blind-mask network in a self-supervised fashion. High amplitude pixels in the training datasets indicate corrupted (i.e., masked) data points. }
  \label{fig:Workflow}
\end{figure}

\subsection{XAI for the mask creation}
\begin{figure}[!htb]
  \centering
  \includegraphics[width=0.9\textwidth]{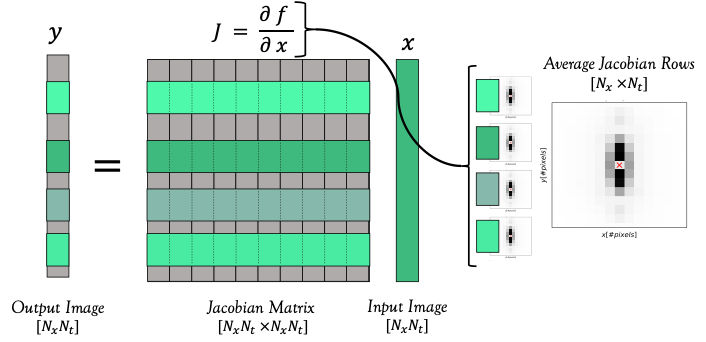}
  \caption{Probing of the neural network's Jacobian Matrix for the creation of the Jacobian Map that illustrates the contributions from input pixels onto a predicted pixel in the output.}
  \label{fig:JacWorkflow}
\end{figure}

The proposed workflow is provided in Figure \ref{fig:Workflow}. In order to determine the contributions from neighbouring pixels, initially a standard blind-spot network is trained, where only the active pixels have their values replaced prior to feeding into the network. After training to convergence, the Jacobian analysis is performed. This analysis is schematically illustrated in Figure \ref{fig:JacWorkflow}, which highlights the linking between the full input image and individual output pixels via rows of the network's Jacobian matrix. The analysis is performed as follows:
\begin{enumerate}
    \item 1000 random points (pixels) are selected from the network's prediction, herein referred to as probe points;
    \item the network's Jacobian matrix is evaluated with respect to the probe points, in other words, 1000 rows of the Jacobian matrix is computed;
    \item the computed Jacobian rows are each reshaped to the dimensions of the input image and a window of $31\times31$ is extracted around the probe points (where this window covers all neighbouring pixels that provide a significant contribution towards the predicted central value);
    \item finally, the absolute values of the 1000 extracted windows are averaged to give a single matrix that represents the average contribution of the neighbouring pixels to any predicted pixels.
\end{enumerate}

\begin{figure}[!htb]
  \centering
  \includegraphics[width=0.9\textwidth]{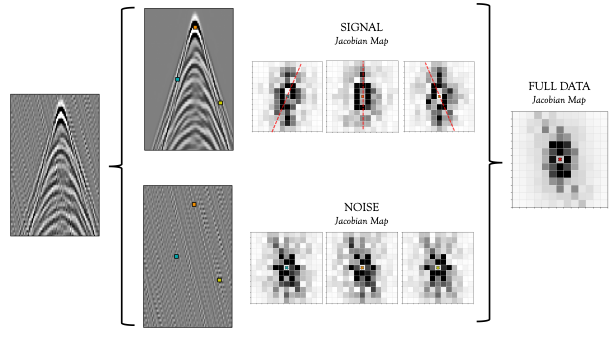}
  \caption{Separate contributions from signal and noise towards the full averaged Jacobian Map. The three annotated points on the single and noise panels correspond to the succeeding Jacobian Maps, paired via their colouring scheme. Red dashed lines on the signal Jacobian Maps indicate the orientation of pixels with a high contribution.}
  \label{fig:JacAss}
\end{figure}

The final averaged matrix is herein referred to as the Jacobian Map. For comparative purposes, we normalise the map such that the contribution from all pixels sums to one. It is important to note that the Jacobian Map represents the total contribution of the pixels without any separation between the pixels that aid the noise recreation and those that are vital for the signal recreation. However, we hypothesise that the signal contributions that the blind-spot network will learn will average out across the probe points, while the contributions from noise oriented in a particular manner will amplify, as schematically illustrated in Figure \ref{fig:JacAss}. 

After the computation of the Jacobian Map, two approaches can be taken to determine the shape of the mask for the subsequent blind-mask network - both of which are investigated in this study. First, the mask can be manually created using the results of the contribution analysis alongside prior knowledge of the pixels that strongly contribute to the signal recreation and the expected noise. This prior knowledge can easily be obtained by training a blind-spot network and computing the Jacobian Map for a clean synthetic dataset with the same frequency content and survey design of the field data of interest - similar to the top row of Figure \ref{fig:JacAss}. We term this approach ``XAI-guided mask design" to emphasise the additional information considered in the mask design process. The alternative approach is to set a contribution limit and all pixels that contribute above such limit in the Jacobian Map become part of the mask. As this incorporates no additional information, we term this ``XAI-driven mask design". The former approach has the benefit that it reduces the removal of pixels that are heavy contributors to the signal prediction, however, its drawback is that it is a manual approach. The latter approach may result in a mask that is larger than necessary and hence removes pixels that provide useful signal information, however, it is an automated approach that only requires the setting of the contribution limit. 

After the mask is selected, the blind-mask training procedure is performed, to teach the network to utilise only unmasked pixels in the prediction of the central pixel. Once the network is trained, the entire raw, noisy data can be passed through the network, without any requirement of pre-processing or masking. The final result is the denoised product arising from the self-supervised, blind-mask scheme.

\section{Numerical Analysis}
To validate our proposed procedure, the approach is first tested on a realistic synthetic dataset contaminated by different types of noise. The dataset has been modelled using a realistic synthetic velocity model that mimics the subsurface structure of the Volve field and a ocean-bottom configuration (see \cite{Ravasi2022} for details). The noise types under investigation include White Gaussian Noise (WGN), time-correlated noise, both isotropic and anisotropic  coloured Gaussian noise (CGN), and pseudo-rig noise. The isotropic CGN has been obtained by filtering a WGN realization with a filter length of 3 in both time and space directions, whilst the anisotropic CGN used a filter length of 4 and 2 in time and space, respectively. The time-correlated noise is generated by applying a trace-wise, band-pass filter to a WGN realisation, and the pseudo-rig noise is generated by creating linear events mimicking waves travelling at a constant speed (i.e., through the sea) convolved with a  non-compact wavelet extending for 10 seconds. 

In all the following examples, the data undergoes \textit{log1p} amplitude scaling, similar to that of \cite{Wang2021}. The benefit of such a scaling method, as opposed to the conventional amplitude gain correction (AGC), is that it can be easily reversed, allowing the amplitudes to return to their original values after the denoising task is complete. In order to obtain an adequate number of training samples, the shots are sliced into overlapping patches of size $64 \times 64$. To ensure patches contain an adequate percentage of signal (as opposed to pre-direct arrival empty contributions), patches whose mean energy is in the $25$th percentile or below are excluded from the training dataset. A 4-layer UNet is trained, with an initial filter size of 32 that doubles per layer. From all of the selected patches, $4096$ are used for training and $1024$ are retained for validation purposes, utilising a batch size of $128$. For both the blind-spot and blind-mask procedures, $2\%$ of the pixels within each input patch are identified as active pixels.

\begin{figure}[!htb]
  \centering
  \includegraphics[width=0.9\textwidth]{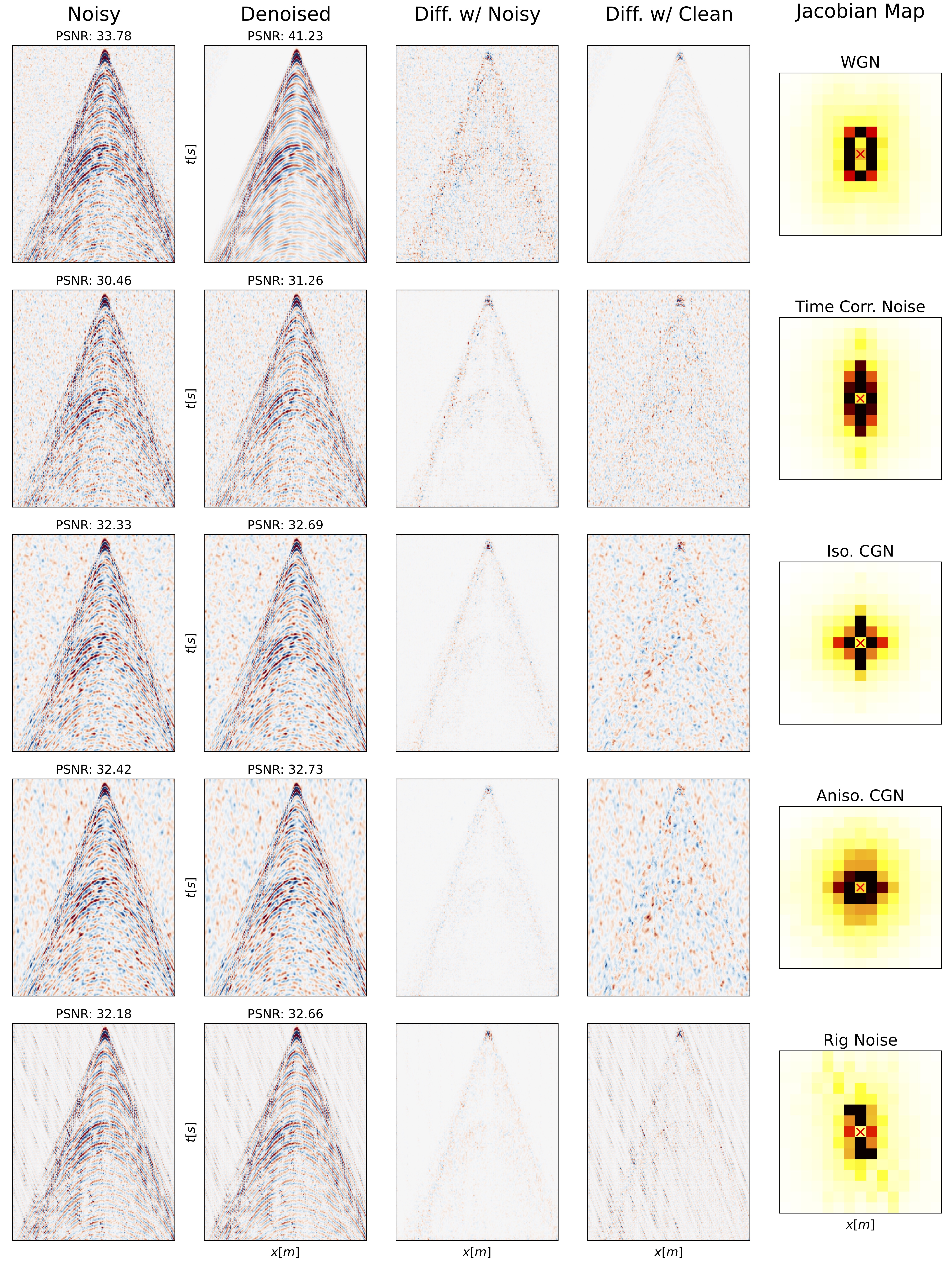}
  \caption{Self-supervised, blind-spot application to synthetic data contaminated with WGN, time-correlated noise, isotropic CGN, anisotropic CGN and pseudo-rig noise, from top to bottom respectively. From left to right, the noisy training data, the `denoised' product, and its difference with the original noisy and clean shots, and finally the averaged Jacobian matrix.}
  \label{fig:Blindspot}
\end{figure}

The application of blind-spot denoising is provided in Figure \ref{fig:Blindspot}, where the first column shows the raw, noisy data and the second column shows the data after blind-spot denoising. The other two columns show the difference between the denoised data with the noisy and clean data, respectively. Whilst the  final column illustrates the averaged Jacobian matrix for 1000 randomly selected pixels in the denoised product. As expected, the network successfully suppresses the WGN in the first example, due to the noise's incoherency between pixels. The Jacobian associated with this result therefore highlights which neighbouring pixels provide a strong contribution towards replicating the signal in the central pixel. Note that due to the powerful denoising capabilities of the blind-spot approach for data contaminated with WGN, this scenario will not be considered any further in this study.

The following four scenarios all contain varying levels of correlated noise and their associated blind-spot products are almost identical to the raw, noisy data. The near-perfect recreation of the noisy data is unsurprising given the coherent nature of the noise components. For the time-correlated noise (second row), the Jacobian analysis clearly indicates a higher contribution from pixels along the same trace as the central pixel. For both CGN examples, the pixels within the filter length of the noise are shown to be high contributors, evident by the rectangular shape of the pixels with a high value in the Jacobian Map. Finally, the Jacobian analysis of the pseudo-rig noise highlights the linear, slightly rotated nature of the noise with stronger contributions of pixels to the top left and bottom right of a central pixel.

To conclude, minimal signal leakage is observed in every example of Figure \ref{fig:Blindspot}, as determined by the lack of coherency in the differences between the predictions and their associated clean shots. This highlights the network's suitability for recreating the complex seismic wavefronts present in such a realistic synthetic dataset.

\subsection{XAI-guided mask design}
\begin{figure}[!htb]
  \centering
  \includegraphics[width=1.\textwidth]{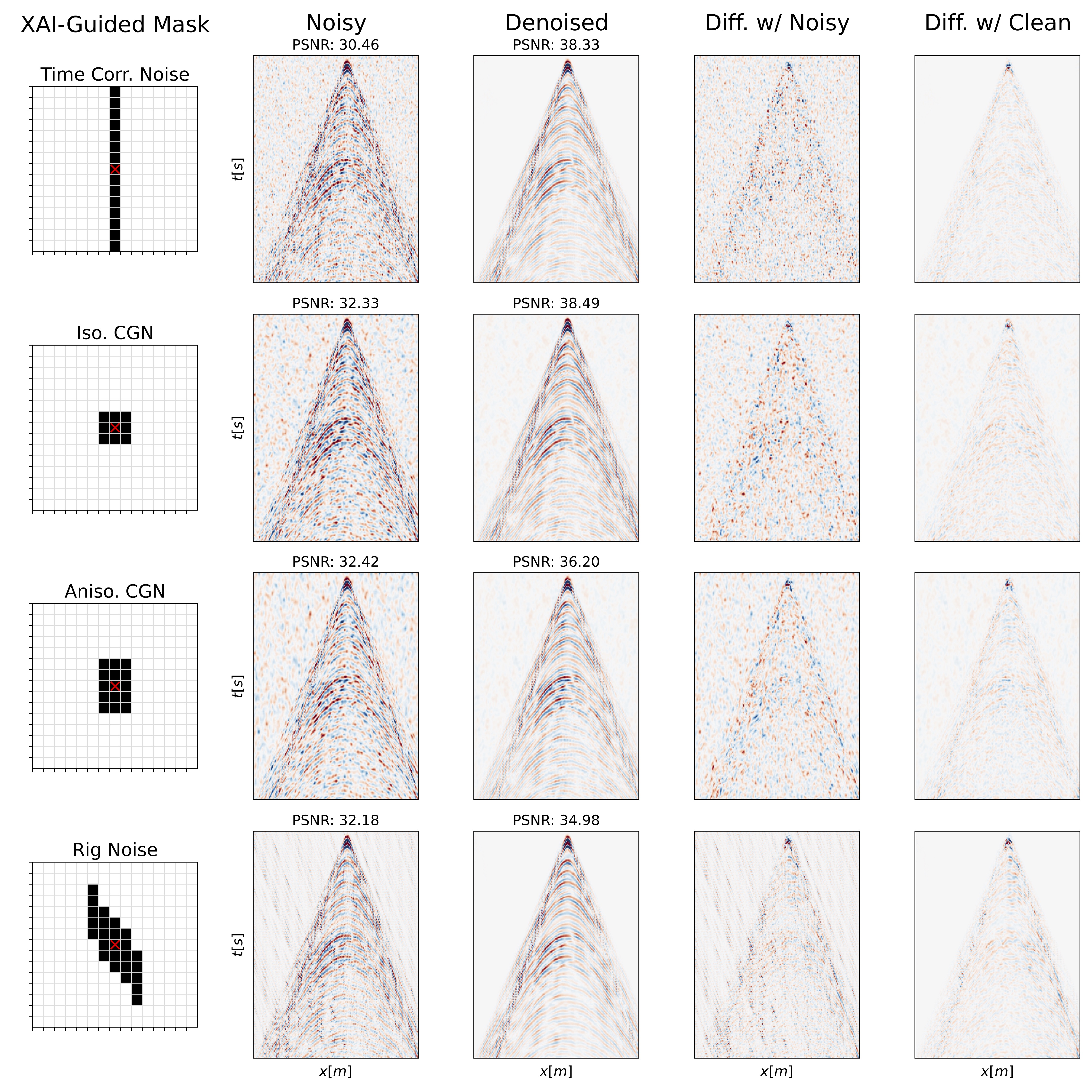}
  \caption{XAI-Guided blind-mask implementation on synthetic data contaminated with time-correlated noise, isotropic CGN, anisotropic CGN and pseudo-rig noise, from top to bottom respectively. The first column represents the XAI-guided noise masks where the red cross marks the active pixel location. The second column shows the noisy data and the third the denoised product. The fourth and fifth columns show the difference between the denoised product and the original noisy data and the clean shot, respectively.}
  \label{fig:ManualBM}
\end{figure}

The XAI-guided methodology allows us to incorporate prior knowledge of the acquisition environment, alongside a modicum of trial-and-error, in order to identify the optimal denoising mask. This allows us to keep highly contributing pixels out of the masked area if we believe them to be likely to provide only signal information. The time-correlated noise is a prime example of this: with the knowledge that the noise is independent between traces and therefore the highly contributing pixels in nearby traces likely provide signal information, then a noise mask can be built only masking highly contributing pixels along the trace direction, as highlighted in the top, left plot of Figure \ref{fig:ManualBM}. The noise masks for the remaining noise scenarios are shown in the left column of Figure \ref{fig:ManualBM} with their associated denoised products displayed in the middle column of the same figure. The two final columns illustrate the difference between the original noisy data and the clean data used for benchmarking. A conventional noise metric, Peak Signal-to-Noise Ratio (PSNR), is provided in the titles for the original noisy data and the denoised products. 

\afterpage{\clearpage}

For all of the considered scenarios, the manually selected masks have resulted in suppression of almost all the noise present in the shot gathers, significantly increasing the PSNR values. However, the noise masks cover a significantly larger area than the previous single blind-spots and therefore some useful signal information is also being masked. This is evident for the bottom two noise masks which are substantial in size and result in a slightly higher level of signal leakage in comparison to the blind-spot approach, when considering the difference between the denoised product and the clean data.

\subsection{XAI-driven mask design}
Unlike the XAI-guided approach, the only variable to be optimised in the XAI-driven approach is the contribution threshold (from the Jacobian analysis) at which a pixel does, or does not, become part of the blind-mask. To ensure consistency across examples, the total contribution of all pixels in the Jacobian map is always provided as a normalised sum; in other words, the sum of the contribution of all pixels within the input image sums to one. The spatial extent of the noise coherency dictates the masks' contribution threshold (due to the number of significantly contributing pixels). In the following analysis, the cut-off percentage has been chosen based on a sweeping analysis. 

\begin{figure}[!htb]
  \centering
  \includegraphics[width=1.\textwidth]{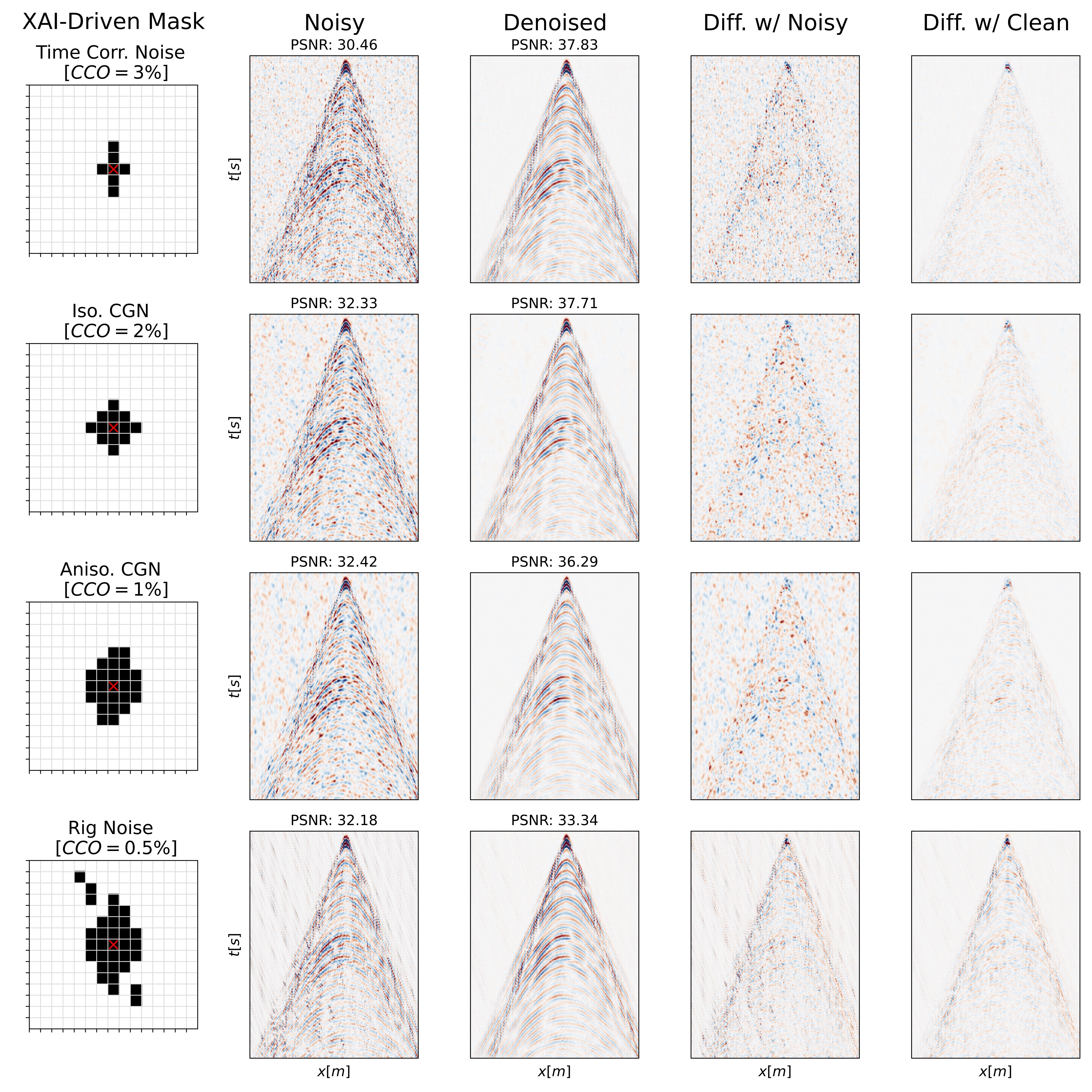}
  \caption{XAI-Driven blind-mask implementation on synthetic data contaminated with time-correlated noise, isotropic CGN, anisotropic CGN and pseudo-rig noise, from top to bottom respectively. The first column represents the automatically selected, XAI-driven noise masks, where the red cross marks the active pixel location and the CCO value represents the contribution cut-off threshold for inclusion in the mask. The second column shows the noisy data and the third the denoised product. The fourth and fifth columns show the difference between the denoised product and the original noisy data and the clean shot, respectively.}
  \label{fig:AutoBM}
\end{figure}

The left column of Figure \ref{fig:AutoBM} illustrates the automatically selected noise masks and their selected thresholds. In general, these masks cover a larger area than their XAI-guided counterparts and their shapes are less obviously rooted in the noise characteristics. For example, the time-correlated noise mask now includes the masking of nearby traces, whilst the pseudo-rig noise mask includes pixels outwith the general masked area. The remainder of Figure \ref{fig:AutoBM} portrays the results of the blind-mask denoising procedure. As in the XAI-guided approach, the XAI-driven procedure has managed to suppress almost all the noise present in each of the four coherent noise scenarios, resulting in significant gains in the PSNR values. Interestingly, despite the masks typically covering a larger area than the XAI-guided approach, a very similar level of signal leakage is visually observed.

As before, the suppression of the pseudo-rig noise poses the biggest challenge to the blind-mask network. As the interfering waves propagate through the same medium as those from the actual survey of interest (i.e., our seismic signal), directionality becomes the only discriminative factor. It is therefore particularly challenging to find a threshold that adequately masks the noise contribution whilst retaining enough information for the network to recreate the desired seismic signal. Whilst the noise has been efficiently suppressed, the high-coherency of the noise and the shared propagation properties between the noise and the signal, has led to higher signal leakage than would be desired. This feature is particularly noticeable in the early arrivals on the right-side of the denoised product (and its corresponding difference plots), where the noise and signal propagate at similar velocities. We conclude that this scenario may be pushing the limits on what can be adequately tackled via self-supervised, blind-mask networks.

\section{Field Results}
\begin{figure}[!htb]
  \centering
  \includegraphics[width=1.\textwidth]{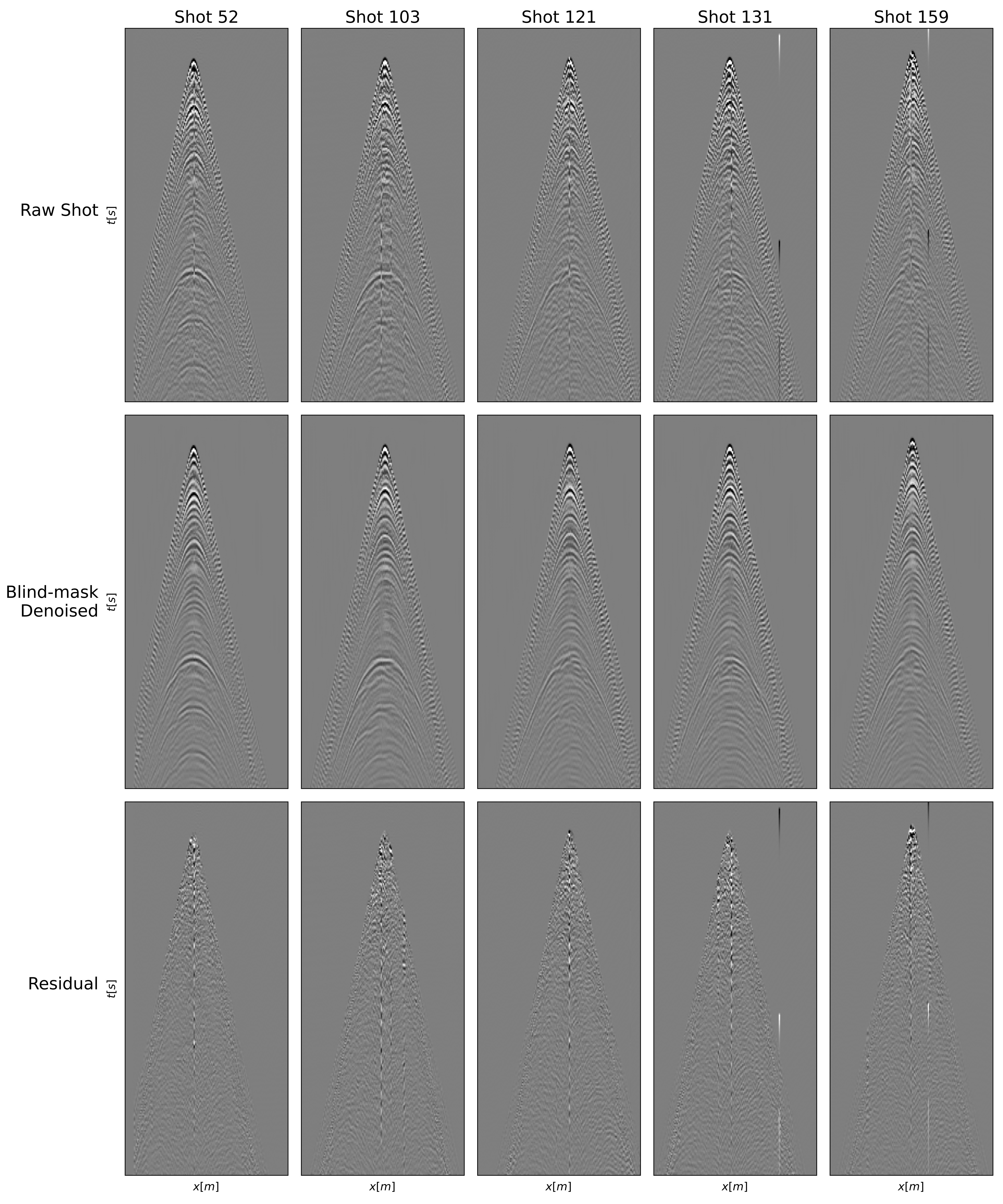}
  \caption{Common shot gathers from the Volve field dataset (top row) and their respective denoised products (middle row) following the XAI-driven masking procedure. The final row illustrates the difference between the raw and denoised shots.}
  \label{fig:FieldVolveMultiShots}
\end{figure}

Following on from the previous examples, which utilised a synthetic dataset inspired from the Volve subsurface, in this section we apply the same methodology to shot gathers from a survey collected over the Volve oil field, previously released by Equinor~\footnote{https://www.equinor.com/energy/volve-data-sharing}. Figure \ref{fig:FieldVolveMultiShots} displays five shot gathers (top row); we observe how such gathers are  contaminated with a small volume of trace-wise noise. As with the synthetic examples, the data undergoes \textit{log1p} amplitude scaling prior to being sliced into overlapping patches of size $64 \times 64$, with patches whose mean energy is below the $15$th percentile being excluded from the training dataset. As before, a 4-layer UNet is trained, with an initial filter size of 32 that doubles per layer. All remaining parameters (e.g., training sample size, number of active pixels, batch size, etc.) are kept consistent with the earlier synthetic studies. To avoid utilising synthetics to aid the mask design, or performing extensive analysis of the field noise, only the automatic, XAI-Driven approach is implemented for the field data, with any pixel contributing more than 2\% to the Jacobian Map being included into the blind-mask.

\begin{figure}[!htb]
  \centering
  \includegraphics[width=1.\textwidth]{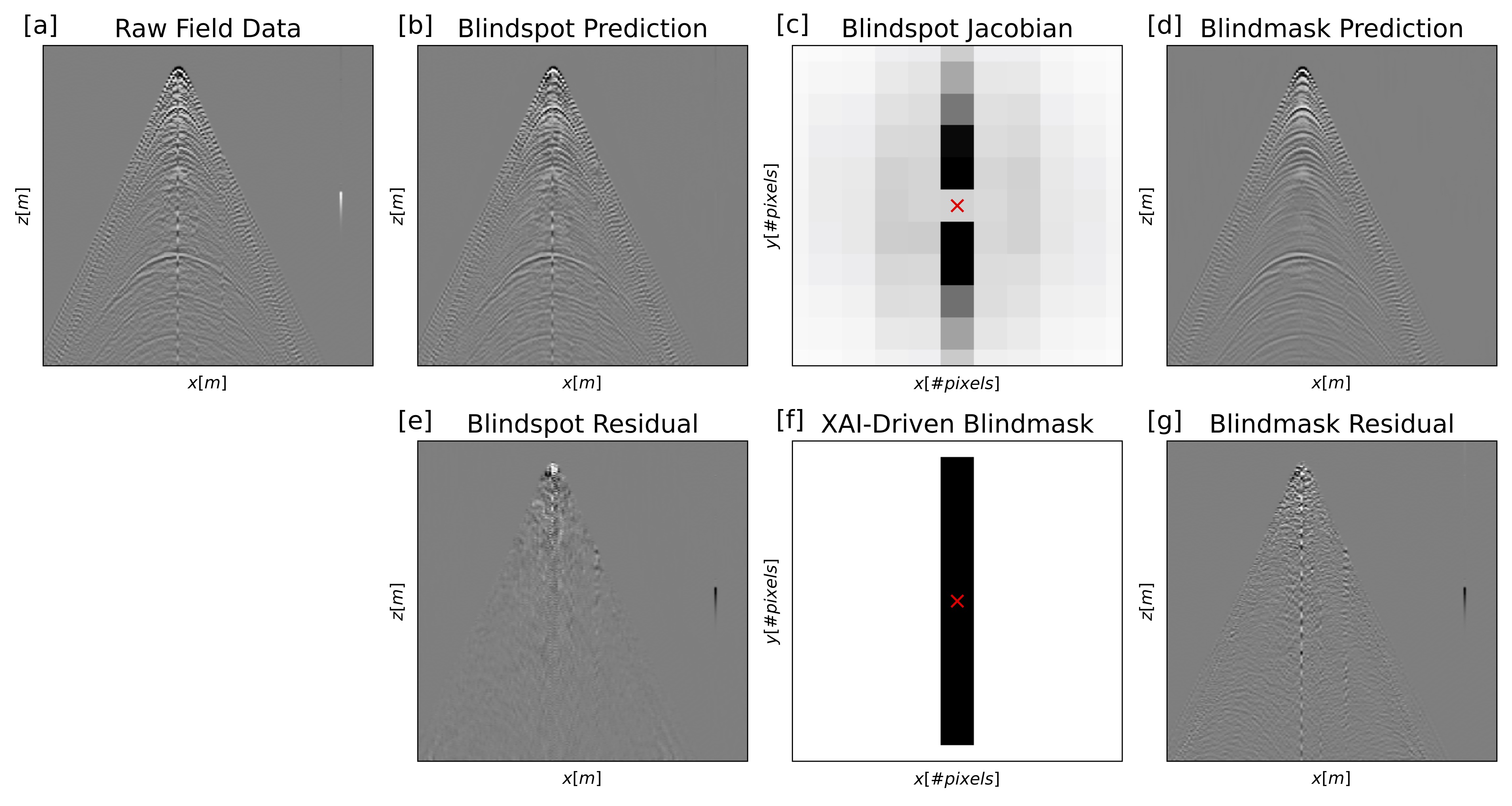}
  \caption{A single shot example of the XAI-driven denoising procedure for the Volve field dataset. (a) the raw data, (b) after blind-spot denoising, and (d) after blind-mask denoising. (c) is the averaged Jacobian matrix and (f) the respective XAI-Driven noise mask. (e) and (g) are the residuals for the blind-spot and blind-mask procedures, respectively.}
  \label{fig:FieldVolve}
\end{figure}

Figure \ref{fig:FieldVolve} details the steps of the mask selection and application with panel b illustrating the denoising performance of the blind-spot network and its associated pixel contributions as defined by the Jacobian Map (panel c). The mask (panel f) is defined as anywhere where the pixel contribution is greater than $2\%$ of the total pixel contributions. Finally, panels d and g illustrate the blind-mask denoising performance and the associated residual. Similarly, the middle row of Figure \ref{fig:FieldVolveMultiShots} illustrates the blind-mask denoising performance on the five shot gathers selected earlier, with their associated residuals (i.e., suppressed noise) displayed in the final row. Whilst the original shots are relatively clean, the computed linear mask has successfully suppressed the trace-wise noise that appeared at different traces across all of the shot gathers. Considering the residual plots, very little coherent energy is observed indicating that minimal signal damage has occurred.

\subsection{Application to post-stack data}
Until now, our proposed methodology has been applied on seismic data prior to migration (i.e., shot-gathers); in other words, denoising is applied to data at the early stages of the seismic processing chain. This is typically where the majority of geophysicists focus their denoising efforts - removing noise before it has a chance to affect down-the-line procedures. However, the proposed methodology is not restricted to any specific stage of processing. To illustrate this, the same methodology is applied to a post-stack time-migrated image from a land field dataset. Similar to the previous examples, a UNet is used; however due to the reduced density of signals in post-stack data, the network is made up of only 2 levels prior to the bottle neck. As only one 2D image is available, the training data is created by dividing it into $7470$ patches of size $32 \times 32$, with an overlap of $4$ samples in the $x-$axis and $6$ samples in the depth-axis.

\begin{figure}[!htb]
  \centering
  \includegraphics[width=1.\textwidth]{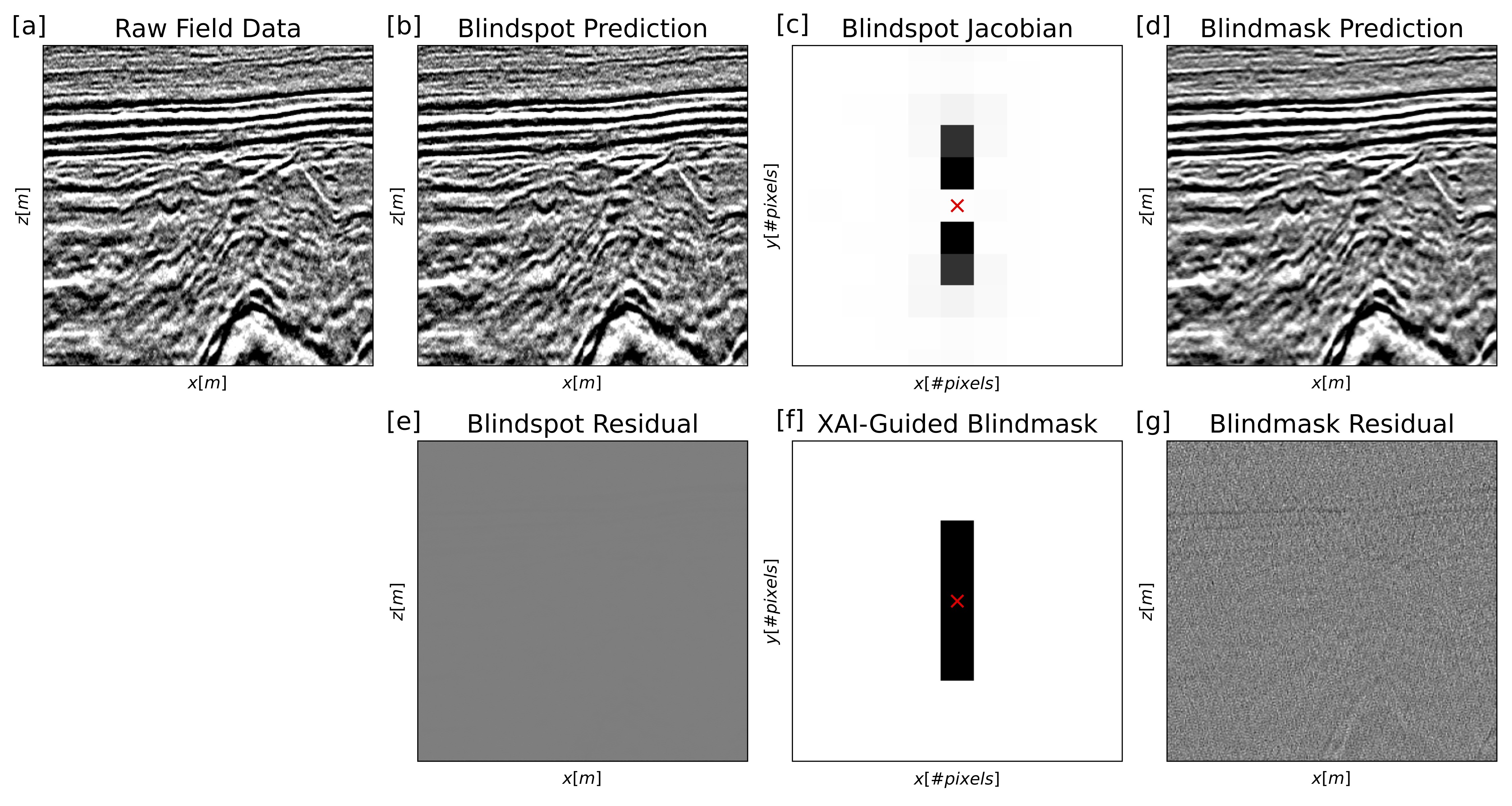}
  \caption{Application of the XAI-driven denoising procedure on a 2D slice of land data after post-stack time migration. (a) the raw data, (b) after blind-spot denoising, and (d) after blind-mask denoising. (c) is the averaged Jacobian matrix and (f) the respective XAI-driven noise mask. (e) and (g) are the residuals for the blind-spot and blind-mask procedures, respectively.}
  \label{fig:FieldChina}
\end{figure}

Figure \ref{fig:FieldChina}(b) illustrates the results of the blind-spot denoising with 2\% of the patch pixels being selected as active pixels and after training the network over 250 epochs. The accompanying Jacobian Map, averaged over all the pixels on inference, is provided in Figure \ref{fig:FieldChina}(c). The subsequent denoising mask is selected by identifying any pixels that contribute more than $2\%$, resulting in the mask portrayed in Figure \ref{fig:FieldChina}(f). A significant volume of noise is removed, as evident in the residual Figure \ref{fig:FieldChina}(g), however an observable level of signal leakage is now present with the linear interfaces in the upper section visible within the residual plot.

\section{Discussion}

\subsection{Denoising}
\begin{figure}[!htb]
  \centering
  \includegraphics[width=1.\textwidth]{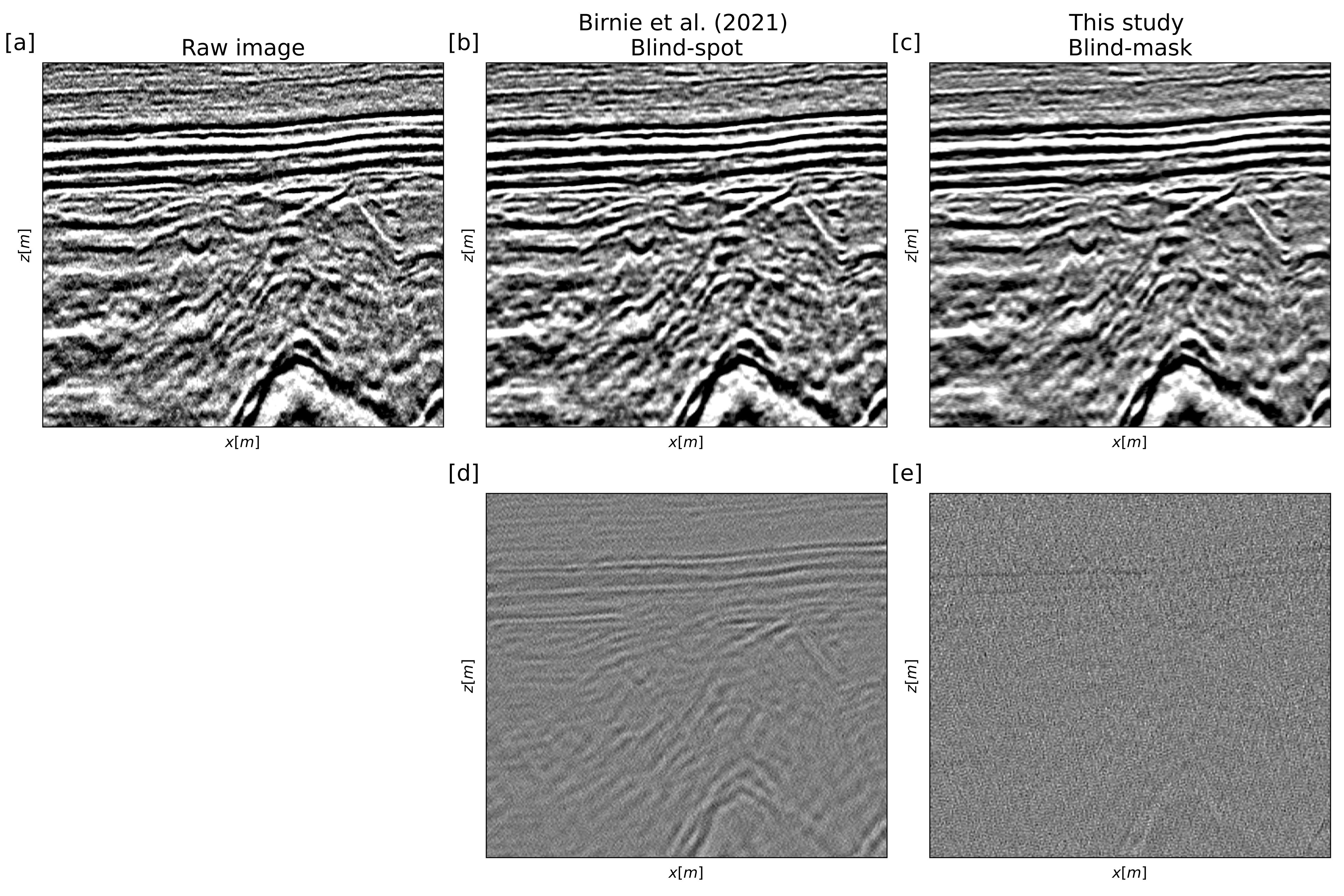}
  \caption{Comparison of the denoising performance of the blind-spot procedure of (b) \cite{Birnie2021} and (c) this study, and (d,e) their associated residuals, respectively. (a) Original, noisy post-stack field dataset.}
  \label{fig:FieldChinaComp}
\end{figure}

In the original implementation of N2V, \cite{Krull2019} selected between 0.5 and $2\%$ of the pixels within a training patch to become active pixels in the training loss. For the seismic denoising scenario, where `random' noise still exhibits loose coherency with neighbouring points, \cite{Birnie2021} illustrated how the assumption of independent and identically distributed noise could be loosened by significantly increasing  the number of active pixels (which will provide a highly corrupted input to the network), alongside adopting an early stopping strategy in the network's training process. In this study, instead of trying to loosen the requirement of noise being random (or corrupted to appear as so), we explicitly design a mask that ensures neighbouring pixels cannot contribute to a central pixel's predicted value. As such, we follow the original approach of \cite{Krull2019} where only a small percentage of the patch is corrupted during the training pre-processing and we allow the training process to run to convergence. One of the benefits of this approach is that the network is being allowed as much time as is needed to fully learn to replicate the seismic signal, without much concern being afforded to the network potentially learning to replicate the coherent noise elements of the data. Figure \ref{fig:FieldChinaComp} provides a comparison for the post-stack field data, using the initial proposal of \cite{Birnie2021} where $1/3$ of the total number of pixels are selected as active pixels and the network is trained for only 15 epochs. By selecting an appropriate blinding mask, as opposed to excessive corruption of the training input, the signal leakage has almost been completely eliminated, whilst the overall volume of noise suppressed has actually increased. Both examples have the exact same network architecture, training samples and network hyperparameters.

Unlike the alternative blind-spot implementation of \cite{Laine2019} - and utilised for seismic deblending by \cite{Luiken2022} - where the receptive field of the neural network is explicitly restricted to remove the central pixel, the pre-processing approach of \cite{Krull2019} assumes that the replaced central pixel value is never useful and therefore the network learns implicitly not to use it. Similarly, the extension to a blind-mask assumes that the network learns not to use the masked pixels. Throughout, the Jacobian has been used to analyse the contribution of the neighbouring pixels on a central pixel's prediction. Therefore, it can be similarly used to evaluate the efficiency of the masking procedure. Table \ref{tab:masked_pix_contr} details the influence of the active pixel and associated mask area as a percentage of the total influence the pixels within the receptive field have on a central pixel's predicted value. In all scenarios the active pixel does provide a small contribution towards it's own predicted value; thereby, illustrating the network hasn't fully learnt to ignore the influence of the active pixel, despite it being corrupted during the training process. For the masked areas, considering the total contribution, the masked pixels' contributions are significantly large, with many contributing more than $10\%$ towards the predicted value. However, the average contribution per masked pixel remains below $1\%$ per masked pixel - lower than some of the contributions of the blind-spot in the standard blind-spot denoising. Whilst this analysis highlights that the network has not fully learnt to ignore the masked area as had been hoped, the influence of these pixels does not have a meaningful value due to their true amplitudes being corrupted (i.e., replaced) during the training. Therefore, it can be concluded that these pixels' contributions only serve to introduce noise into the prediction procedure. Given the promising denoising results, future work will investigate the level to which the mask efficiency is required in order to achieve the best possible denoised product.

\begin{table}[!htb]
\centering
\caption{Contribution of masked pixels for the blind-spot and blind-mask denoising schemes, given as a percentage of the total overall pixel contributions within the Jacobian Map. }
\label{tab:masked_pix_contr}
\begin{tabular}{lllcccc}
\multicolumn{3}{l}{} &  & \multicolumn{2}{c}{\textbf{CGN}} &  \\
\multicolumn{3}{l}{\multirow{-2}{*}{}} & \multirow{-2}{*}{\textbf{Time Corr.}} & \multicolumn{1}{l}{\textbf{Iso.}} & \multicolumn{1}{l}{\textbf{Aniso.}} & \multirow{-2}{*}{\textbf{Rig noise}} \\ \hline
\multicolumn{1}{r}{} & \multicolumn{2}{c}{\textit{Active Pixel}} & 0.8 & 1.2 & 0.9 & 0.5 \\
\multicolumn{1}{r}{\multirow{-2}{*}{\textbf{Blind-spot}}} & \multicolumn{2}{c}{\cellcolor[HTML]{C0C0C0}\textit{Masked Area}} & \cellcolor[HTML]{C0C0C0} & \cellcolor[HTML]{C0C0C0} & \cellcolor[HTML]{C0C0C0} & \cellcolor[HTML]{C0C0C0} \\ \hline
\multicolumn{1}{r}{} & \multicolumn{2}{c}{\textit{Active Pixel}} & 1.6 & 0.7 & 0.7 & 0.7 \\
\multicolumn{1}{r}{} &  & \textit{Total} & 13.2 & 6.0 & 11.2 & 14.8 \\
\multicolumn{1}{r}{\multirow{-3}{*}{\textbf{\begin{tabular}[c]{@{}r@{}}XAI-Guided\\  Blind-mask\end{tabular}}}} & \multirow{-2}{*}{\textit{Masked Area}} & \textit{Avg./Pixel} & 0.4 & 0.7 & 0.7 & 0.6 \\ \hline
 & \multicolumn{2}{c}{\textit{Active Pixel}} & 1.0 & 0.8 & 0.5 & 0.5 \\
 &  & \textit{Total} & 6.0 & 8.8 & 12.9 & 16.2 \\
\multirow{-3}{*}{\textbf{\begin{tabular}[c]{@{}l@{}}XAI-Driven \\ Blind-mask\end{tabular}}} & \multirow{-2}{*}{\textit{Masked Area}} & \textit{Avg./Pixel} & 0.9 & 0.7 & 0.5 & 0.5 \\ \hline
\end{tabular}
\end{table}

\subsection{Jacobian Analysis}
One of the challenges with utilising the Jacobian matrix for self-supervised, blind-spot denoising is that there is no clear delineation between pixels that contribute towards the noise component of the central pixel versus those that contribute towards the signal component. In our synthetic analysis, we proposed the use of an XAI-guided procedure where the researcher could utilize their prior knowledge of the pixel contributions as a complement to the output of the Jacobian matrix analysis. Such prior knowledge could be knowledge of the noise statistics helping the researcher to identify any highly contributing pixels that may be providing only signal information. Alternatively, it could be knowledge from a prior synthetic investigation which identified the pixels that are likely to contribute signal information. Whilst the XAI-guided approach allows the incorporation of prior knowledge as a means to reduce the masking of pixels that provide useful information on the seismic signal, this introduces some manual labour in the process. To the contrary, the XAI-driven methodology is more heavy-handed with respect to the selection of the mask, however it is a fully automated procedure that requires no prior knowledge on the monitoring conditions or the properties of the seismic signals. As the mask design in the XAI-guided approach is subjective, a direct comparison between XAI-guided and XAI-driven is not possible. However, in the numerical experiment section of this paper we show that both are suitable to the synthetic case. Due to a lack of prior knowledge, only the XAI-driven approach was implemented for the field data studies.

\section{Conclusions}
Self-supervised, blind-spot networks have been shown to be powerful denoisers for seismic data contaminated by random noise; however, they require an additional masking step for the suppression of coherent noise. The mask design is often non-trivial, requiring extensive knowledge of the noise characteristics and how the network will discriminate between them and the signal that we wish to preserve. In this study, we have proposed the use of bias-free networks, which allow the denoising process to be interpreted as a linear combination of the input pixels, where the weights can be identified by computing the Jacobian matrix of the network itself; ultimately allowing the identification of pixels with a strong influence on any predicted value. This knowledge is utilised to automatically design noise masks to hide said pixels during the denoising networks training process. The resulting blind-mask networks are shown to successfully suppress a range of coherent noise types on realistic synthetic seismic shot gathers. Furthermore, the methodology is applied to two field datasets at different ends of the processing pipeline - common shot gathers and post-stack images. The successful suppression of noise in the field datasets highlights not only the robust, denoising capability of the proposed procedure, but also its general applicability to a variety seismic denoising applications. To conclude, the proposed procedure offers an automated, deep learning-based workflow that operates directly on the raw field data with no requirement of prior knowledge on the site conditions. 

\section{Acknowledgements}
The authors thank the KAUST Deep Imaging Group and Seismic Wave Analysis Group for insightful discussions, and in particular, Professor Tariq Alkhalifah, Dr Nick Luiken, and Ms Sixiu Liu. For computer time, this research used the resources of the Supercomputing Laboratory at King Abdullah University of Science \& Technology (KAUST) in Thuwal, Saudi Arabia. CB kindly acknowledges funding from Saudi Aramco for aspects of this work.

\section{Data Availability Statement}
All data used in this study are openly available within the public domain. The Volve synthetic waveform data is available on the Zenodo data platform, at \url{http://doi.org/10.5281/zenodo.6572286}. The Volve field data is available directly from Equinor ASA at \url{https://www.equinor.com/energy/volve-data-sharing}. The post-stack image is available in the Madagascar data repository, at \url{https://github.com/ahay/src}. The code to generate the synthetic noise instances is available from the authors by request.

\typeout{}
\bibliographystyle{unsrt} 
\bibliography{main}

\newpage

\end{document}